\documentclass[aps,plb,a4paper,twocolumn,showpacs,groupedaddress,nofootinbib]{revtex4}
\usepackage{amsmath}
\usepackage{amssymb}
\usepackage{graphicx}
\usepackage{color}
\usepackage{geometry}
\begin{document}

\title{Entropy Density and Speed of Sound from Improved Energy-Momentum Tensor in Lattice QCD}
\author{Mushtaq  Loan$^{a}$\footnote{Corresponding author} and Nasser Demir$^{b}$}
\affiliation{$^{a}$Centre for Medical Radiation Physics, School of Physics, University of Wollongong, NSW 2522,  Australia\\
$^{b}$Department of Physics. Kuwait University, Safat 13060,  Kuwait}
\date{\today}

\begin{abstract}
 We present a lattice calculation of the entropy density $s/T^{3}$  and speed of sound  $c_{s}^{2}$ of gluedynamics near the critical temperature, $T_{c}$,  
 in the deconfined phase.  By exploring the temperature dependence of entropy density in this region, we aim to analyse the significant discrepancies between 
 the previous computations.  The calculation of entropy density is carried out by numerical simulations of  $O(a^{4})$ mean-field improved energy-momentum 
 tensor (EMT) of SU(3) gauge theory on the lattice.  We expand on reaching $O(a^{4})$ improvement using tadpole-improved Symanzik action. The entropy 
 density is calculated directly from the expectation value of the space-time component of the improved EMT in the presence of shifted boundary conditions at 
 several lattice spacings ($a \approx  0.043 - 0.012$ fm).   The absence of ultraviolet divergences and the minimal finite-size effects allow for the precision 
 determination of the entropy density and its extrapolation to the continuum limit. As expected, the resulting entropy density displays the expected behaviour 
 of rapid increase near the critical temperature in the deconfined phase followed by a slow increase in $2T_{c}\leq T\leq 3T_{c}$ region, suggesting a 
 logarithmic dependence on the temperature.  A quantitative comparison of $s/T^{3}$  shows good agreement with Pade approximation and lattice results of 
 previous high-precision data obtained using the gradient flow method. We observe that at temperatures of about $3T_{c}$,  deviations of entropy density 
 from the Stefan-Boltzmann value for a free theory are about 10$\%$. It is shown that the speed of sound in SU(3) gluedynamics is found to be $c_{s}^{2}
 \leq 0.333$ in the temperature region $1.06T_{c}\leq T\leq 3.05T_{c}$ explored in this study. The results are found to agree with the corresponding analytic and numerical estimates.

\end{abstract}
\pacs{05.07.Ce, 11.10.Wx, 11.15.Ha, 12.38.Mh}

\maketitle

\section{Introduction}
For decades, quantum chromodynamics (QCD) at finite temperature and density has been the focus of intense theoretical research \cite{Sayantan2021} (and 
references therein). A comprehensive understanding of thermodynamic observables in QCD, such as energy density and pressure, is of vital interest in 
particle physics and cosmology. Apart from the obvious interest, the collective behaviour of these quantities provides essential information for studying the 
evolution of the universe in its early stages. In particular, topological susceptibility in QCD, which provides the theoretical input for axion cosmology, has received considerable recent attention.  Relativistic heavy-ion colliders are now being used to reproduce and study the equilibrium and transport 
properties of strongly-interacting matter, where the equation of state (EoS) is a crucial component of data analysis 
\cite{Bernhard2016,Parotto2020,Monnai2019,Everett2020,Derradi2016}. Recent relativistic heavy ion collision experiments at RHIC and LHC aimed at 
studying the properties of strongly interacting matter under extreme conditions have revealed various important properties of QCD. 
\cite{Molnar2002,Gyu2005,Shuryak2005,Peshier2005}. As a result, energy density, pressure, and entropy density have dominated this discussion  \cite{Heinz2013,Oll2008,Gale2013}. 

Lattice Quantum Chromodynamics (QCD)  provides an excellent tool for studying the high-temperature behaviour of QCD. Lattice simulation results for the EoS play a critical role in exploring the QCD phase diagram in the $T - \mu_{B}$ plane. Using the conventional integral method of calculating the trace anomaly, Boyd \emph{et al.} performed the first lattice measurement of the EoS up to $T \sim 5T_{c}$  \cite{Boyd1996}. This study was extended by several other extensive lattice studies, most of which used a modified version of the integral method to obtain the thermodynamic observables for SU(3) pure gauge theory \cite{Okamoto2001,Umeda2009,Borsanyi2012,Giusti2016} and in full QCD \cite{Borsanyi2014,Bazavov2014}  in a broader temperature range up to $\sim 1000T_{c}$ with higher accuracy. However, besides the requirement of subtraction of ultraviolet divergences in the integral method, these calculations show significant discrepancies in the region near the critical temperature \cite{Philpsen2013,Ding2015}. Even though the differences in trace anomalies at two different temperatures cancel these ultraviolet divergences, conducting simulations at two different temperature scales tuned to the same bare parameters is computationally challenging. Such difficulties have restricted the investigation of the continuum limit of thermodynamic observables only up to temperatures of about 1-2 GeV \cite{Borsanyi2014,Bazavov2014,Bazavov2018}.  Recently, the equation of state at finite temperature has been studied by directly measuring renormalised energy-momentum tensor constructed from the flowed field at nonzero flow-time using the gradient flow method \cite{Suzuki2013,Asakawa2015,Luscher2010,Narayanan2006,Fodor2012}. It was observed that the signal-to-noise ratio in the energy-momentum tensor could be significantly enhanced by suppressing the ultraviolet modes \cite{Luscher2010,Luscher2011}.  The technique has been successfully extended to full QCD \cite{Makino2014,Itou2016,Taniguchi} and appears to be a viable method to study the correlation functions and transport coefficients of the quark-gluon plasma.

Giusti and Pepe proposed the strategy for determining the equation of state of a relativistic thermal quantum theory by defining the 
ensemble in a moving reference frame \cite{Giusti2011,Giusti2011b,Giusti2013,Giusti2013M,Giusti2011M,Giusti2017}. This allowed 
the calculation of entropy density directly from the off-diagonal components of the  EM. The other thermal variables were calculated 
from the entropy function through thermodynamic identities. The results of their studies show a good agreement with the earlier 
calculations \cite{Borsanyi2012} for high temperatures but differ on a $4\%$ level below $3T_{c}$. Despite the impressive progress 
over the last few years, uncertainties in the thermodynamics quantities are still relatively significant. Here we present lattice 
calculation of an improved energy-moment tensor in the Euclidean SU(3) gauge theory to extract the entropy density and speed of 
sound. Developing on the technique of ensemble in a moving reference system, we aim to analyse the discrepancy in entropy 
density near $T_{c}$ by using a  tadpole-improved energy-momentum tensor to compute the matrix elements of  EMT under shifted 
boundary conditions.   Another observable of interest is the speed of sound, a vital input characterising different phases.  If 
QGP at high temperatures was qualitatively close to an ideal gas of non-interactive massless particles, then the speed of sound 
would approach the Stefan-Boltzmann limit.

\section{Method}

\subsection{Tadpole-improved Energy-Momentum Tensor}

The thermal theory is defined on a finite four-dimensional lattice of spatial volume $L_{s}^{3}$, temporal direction $L_{0}$, and 
lattice spacing $a$. The gauge field satisfies periodic boundary conditions in the spatial directions and shifted boundary conditions in 
the temporal direction
\begin{equation}
U_{\mu}(L_{0}, \pmb{x}) = U_{\mu}(0, \pmb{x} - L_{0}\pmb{\xi}),
\label{eqn1}
\end{equation}
where $U(L_{0}, {\bf x})$ are the link variables and $\pmb\xi\in \mathbb{R}^{3}$ is the shift vector in the temporal direction and 
corresponds to the Euclidean velocity of the moving frame  \cite{Giusti2017}. The periodic boundary conditions are restored in the 
rest frame,  $\pmb{\xi}=0$.  In the presence of a mass gap at the zero-temperature limit of the theory, the invariance of  
the theory under the Poincar\'e group forces its free energy to be independent of the shift $\pmb{\xi}$. At nonzero temperatures, the 
finite time-length $L_{0}$ breaks the Lorentz group softly; consequently, the free energy depends on the shift explicitly but only 
through the inverse temperature $L_{0}\sqrt{1+\pmb{\xi}^{2}}$.

\begin{figure}[!h]
\scalebox{0.55}{\includegraphics{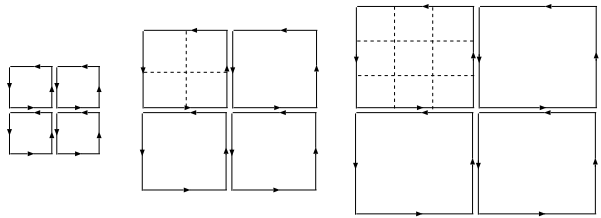}} 
\caption{\label{figa} The $1\times 1$, $2\times 2$, and $3\times 3$ loops are used to construct the clover term for the improved energy-momentum tensor.}
\end{figure}

The energy-momentum tensor of the gauge field theory has the form:
\begin{equation}
T_{\mu\nu}(x) = \frac{1}{g_{0}^{2}}\left[F_{\mu\alpha}^{a} - \frac{1}{4}\delta_{\mu\nu}F_{\alpha\beta}^{a}F_{\alpha\beta}^{a}\right],
\label{eqn2}
\end{equation}
where the gluon field strength tensor is defined as
\begin{equation}
F_{\mu\nu}^{a}(x) = -\frac{i}{4a^{2}}\left[\bigg(Q_{\mu\nu}(x) - Q_{\mu\nu}^{\dagger}(x)\bigg)T^{a}\right],
\label{eqn2b}
\end{equation}
and
\begin{equation}
Q_{\mu\nu} = \frac{1}{4}\bigg(U_{\mu\nu}(x)+U_{-\nu\mu}(x)+U_{\nu-\mu}(x)+U_{-\mu-\nu}(x)\bigg)
\label{eqn3}
\end{equation}
is the sum of the four plaquette terms. The gluon field strength tensor has $O(a^{2})$ discretisation errors. To improve the discretisation of the gluon field strength tensor, we incorporate additional higher "clover" loops (Fig. \ref{figa}) in $F_{\mu\nu}$. In general, we define the following  improved field strength tensor 
\begin{eqnarray}
F_{\mu\nu}^{imp}(x) &= & k_{1}F_{\mu\nu}^{1\times 1}+ k_{2}F_{\mu\nu}^{2\times 2}+\frac{k_{3}}{2}\bigg(F_{\mu\nu}^{2\times 1}+
F_{\mu\nu}^{1\times 2}\bigg)\nonumber\\
& & \frac{ k_{4}}{2}\bigg(F_{\mu\nu}^{3\times 1}+ F_{\mu\nu}^{1\times 3}\bigg) + k_{5}F_{\mu\nu}^{3\times 3},
\label{eqn4}
\end{eqnarray}
where
\begin{eqnarray}
F_{\mu\nu}^{m\times n}(x) = - \frac{i}{4a^{2}}\left[\bigg(Q_{\mu\nu}^{m\times n}(x) - Q_{\mu\nu}^{\dagger m\times n}(x)\bigg)T^{a}\right]
\label{eqn5}
\end{eqnarray}
$k_{i}$ are the constant coefficients and $Q_{\mu\nu}^{m\times n}(x)$ corresponds to the sum of the four $m\times n$ loops in the clover formation. 

For computational efficiency, we consider a 3-loop improved field strength tensor ($k_{4}=k_{5}=0$) with mean field improved coefficients,
\begin{equation}
F_{\mu\nu}^{3L}(x) = \frac{3}{2u_{0}^{4}}F_{\mu\nu}^{1\times 1}(x)
-\frac{3}{20u_{0}^{8}}F_{\mu\nu}^{2\times 2}(x)+\frac{1}{90u_{0}^{12}}F_{\mu\nu}^{3\times 3}(x)
\label{eqn6}
\end{equation}
The improved energy-momentum tensor is then represented by
\begin{equation}
T_{\mu\nu}^{Imp}(x) = \frac{1}{g_{0}^{2}}\left[F_{\mu\alpha}^{3L} - \frac{1}{4}\delta_{\mu\nu}F_{\alpha\beta}^{3L}F_{\alpha\beta}^{3L}\right].
\label{eqn6b}
\end{equation}
We are interested in the off-diagonal components of $T_{\mu\nu}^{Imp}$ 
\begin{equation}
\langle T_{0k}^{Imp} \rangle_{\pmb{\xi}}= \frac{\xi_{k}}{1-\xi^{2}_{k}}\left[\langle T_{00}^{imp} \rangle_{\pmb{\xi}} - \langle T_{kk}^{Imp} \rangle_{\pmb{\xi}}\right], 
\label{eqn7}
\end{equation}
where $\langle \cdot \rangle_{\pmb{\xi}}$ represents the expectation value computed in the thermal system with 
shift $\pmb{\xi}$. 

\subsection{Entropy Density and Speed of Sound}
In a moving reference frame, the entropy density is computed from the momentum density of improved  normalised energy-momentum tensor through the relation:
\begin{equation}
s = - \frac{L_{0}(1+{\xi}^{2})^{3/2}}{\xi_{k}}\langle T_{0k}^{Imp}\rangle_{\xi}, \hspace{0.50cm} \xi_{k} \neq 0.
\label{eqn11}
\end{equation}
Following the convention to express the value of entropy density in terms of entropy-density-to-temperature ratio, $s/T^{3}$, we have from Eq. (\ref{eqn11})
\begin{equation}
s/T^{3}= - \frac{L_{0}^{4}(1+{\xi}^{2})^{3}}{\xi_{k}}\langle T_{0k}^{Imp}\rangle_{\xi}.
\label{eqn13}
\end{equation}

Another quantity of interest in describing the evolution of quark-gluon plasma is the speed of sound. Considering the asymptotic freedom at high temperatures (energies), the QGP can be considered an ideal gas of quarks and gluons. However, near the critical temperature in the deconfined region, the system displays a non-ideal behaviour that is well described by the quasiparticle model.  Using entropy and specific heat, the speed of sound is obtained as \cite{Yagi2005,Khan2006,Aoki2006}
\begin{eqnarray}
c_{s}^{2} &=& \frac{\partial P}{\partial \epsilon} = \frac{\partial\ln T}{\partial\ln s}\nonumber\\
& = & \frac{s}{T} \frac{\partial T}{\partial s}.
\label{eqn13b}
\end{eqnarray}
It has been demonstrated that the speed of sound has a possible discontinuity at $T = T_{c}$ \cite{Borsanyi2012}. The speed of sound can then be written  in the more suggestive form
\begin{equation}
c_{s}^{2} = \frac{s\xi_{k}}{Z_{T}(1+\xi^{2})\left[\langle T_{0k}^{imp.}\rangle - \frac{d}{dT}\langle T_{0k}^{imp.}\rangle\right]}.
\label{eqn13e}
\end{equation}
For a scale-invariant system in three spatial dimensions in SU(3) Yang-Mills theory, the speed of sound should be equal to 1/3. However, due to a violation in conformal symmetry in the confinement-deconfinement region,  $c_{s}^{2}$ is expected to deviate from the Stefam-Boltzmann limit for an ideal gas limit of massless particles. 

\section{Results and Discussion}

In our computation, we opted for mean-field improved Symanzik gauge action \cite{Alford1995}, which has lead $O(\alpha_{s}a^{2})$ and $O(a^{4})$
discretisation errors and gives results close to the continuum on coarse lattices. The expectation values of the energy-momentum tensor are measured on the lattice $64^{3}\times N_{t}$ ($N_{t}= 8, 10, 12, 16, 20$ with $\beta$ values in the range $6.10 - 7.52$, which correspond to the temperatures  $T/T_{c} = 1.06 - 3.05$.  Perturbative studies indicate that for these values of  $N_{t}$, the results should dictate minor discretisation errors for the entropy density \cite{Brida2018}. To investigate the size of finite volume effects in the relevant matrix elements of the energy-momentum tensor, we generate three ensembles over a larger spatial resolution of $N_{s}=128$. Gauge configurations are generated using a mixture of pseudo-heatbath and over-relaxation sweeps. Configurations are given a hot start and  500 compound sweeps to equilibrate. We define a compound as one pseudo-heatbath update sweep and five over-relaxation sweeps. After thermalisation, configurations are stored every 250 compound sweeps to eliminate the autocorrelation.  For each $\beta$ value, 20,000 to 30,000 gauge configurations are stored for the measurements. The measured data are divided into bins; each is considered an independent height for analysis. Errors in Monte Carlo data have been estimated using both the jackknife and binning techniques. The critical temperature is set in the units of the Sommer scale. In the $1.3T_{c} - 2.80T_{c}$, the accuracy of the temperature was observed to be about a percent. The parameters of the simulations at various $\beta$ values are summarised in Tab. \ref{tab1}.
\begin{table}[ht!]
\begin{center}
\caption{Simulation parameters and statistics on a $64^{3}\times N_{t}$  lattice with bin size $N_{size}$, number of bins $N_{b}$, and $\pmb{\xi} =(1,0,0)$ .}
\label{tab1}
\begin{tabular}{ccccccccccc}\hline\hline
$\beta$ & $N_{s}$ & $N_{t}$ & $N_{size}$ & $N_{b}$ & $T/T_{c}$ \\ \hline
6.10   & 64 & 20 & 2000 & 10 & 1.064\\
6.23   & 64 & 20 & 2000 & 10 & 1.123\\
6.41   & 64 & 16 & 2000 & 10 & 1.294\\
6.55   & 64 & 12& 2000 & 10 & 1.513\\
6.69   & 64 & 12& 2000 & 10 & 1.860 \\
6.81   & 64 & 10 & 2000 & 10 & 2.079\\
6.94   & 64 & 10 & 2000 & 10 & 2.491 \\ 
7.18   & 64 & 10 & 2000 & 10 & 2.784\\ 
7.31   & 64 & 8 & 2000 & 15 & 2.932\\
7.52   & 64 & 8 & 2000 & 15 & 3.052 \\ \hline
\end{tabular}
\end{center}
\end{table}

\begin{figure}[ht!]
\scalebox{0.45}{\includegraphics{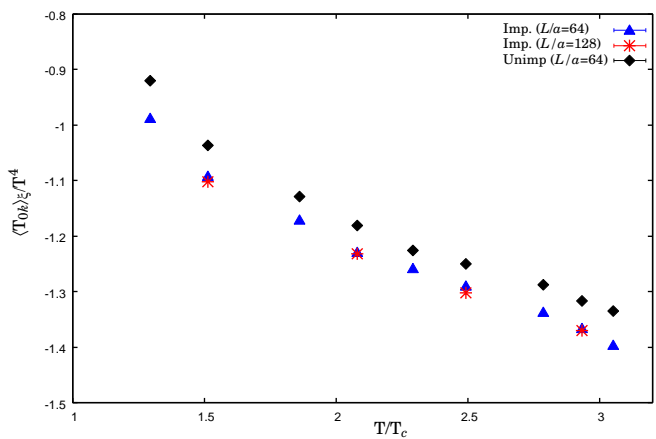}} 
\caption{\label{fig2} Expectation values of the space-time components of bare energy-momentum tensor as a function of $T/T_{c}$  at $\pmb{\xi}=(1,0,0)$. The data are generated on lattices with $N_{s}= 64$ (triangles) and $128$ (stars)  using improved EMT. The data generated using unimproved  EMT (diamonds) is also shown for comparison. The error bars are smaller than the symbols.}
\end{figure}

The expectation values of the bare $T_{0k}$ of the improved and unimproved energy-momentum tensor for the shift value $\pmb{\xi} = (1,0,0)$  for the ensembles considered here are reported in Table \ref{tab2} and displayed in Fig. \ref{fig2}. At fixed $N_{t}$, $\beta$ and the number of measurements, the results obtained on larger spatial volume indicate that the expectation value of momentum flux does not seem to have any strong volume dependence. For $N_{s}=128$, we  typically reach a precision of $\approx 0.1 - 0.4\%$.  The statistical errors in the improved $\langle T_{0k}\rangle$  are smaller than the symbols and grow linearly from 0.10$\%$ to 0.25$\%$. For the comparison, we observe that the results obtained using unimproved bare EMT (represented by the black diamond symbols in Fig. \ref{fig2}), differ by about $5 - 7\%$  for the temperatures investigated in this study.
\begin{table}[ht!]
\begin{center}
\caption{The expectation value of the space-time components of $T_{0k}$  for improved and unimproved (Eq. \ref{eqn3}) definitions of the momentum density of  EMT at $N_{s} = 64$ and $128$, and  $\pmb{\xi}= (1,0,0)$.}
\label{tab2}
\begin{tabular}{ccccccccccc}\hline\hline
$T/T_{c}$ & \multicolumn{2}{c}{$\langle T_{0k}\rangle /T^{4}$} \\  \hline
    &  \multicolumn{2}{c}{Imp}  & Unimp \\
    & ($N_{s}= 64$) & ($N_{s}=128$) & ($N_{s} = 64$) \\ \hline
1.294  &  -0.9878(12) &                        & -0.9196(31)\\
1.513  & -1.0926(11)  & -1.1021(10)    &  -1.0367(48)  \\
1.860  & -1.1709(14) &                         &   -1.1282(39)   \\
2.079  &  -1.2294(13) & -1.234(12)     &  -1.1805(44)  \\ 
2.291  & -1.2584(16) &                        &  -1.2254(37)  \\ 
2.491  & -1.2906(18) &-1.3021(20)    &  -1.2497(47)  \\
2.784  & -1.3378(21) &                       &  -1.2874(35)   \\
2.932  & -1.3661(17) & -1.3698(16)   &  -1.3171(43)   \\
3.051  & -1.3965(23) &                      &   -1.3352(42)  \\  \hline 
\end{tabular}
\end{center}
\end{table}
Given that the mean-field improved field strength tensor and the energy-momentum tensor used in the simulations lead to deviations of $O(a^{4})$ of the lattice Lagrangian from its continuum counterpart, the thermodynamic observables like energy density, pressure, and entropy density calculated using tensors will then deviate from the continuum values by $O((aT)^{4})$ terms.   Based on the analysis of the entropy density on various lattice sizes, we attempt to extrapolate $s/T^{3}$ to the continuum limit using
\begin{equation}
\bigg(s/T^{3}\bigg)_{a} = \bigg(s/T^{3}\bigg)_{0}+c_{1}/N_{t}^{2}.
\label{eqnf1}
\end{equation}

\begin{figure}[ht!]
\scalebox{0.45}{\includegraphics{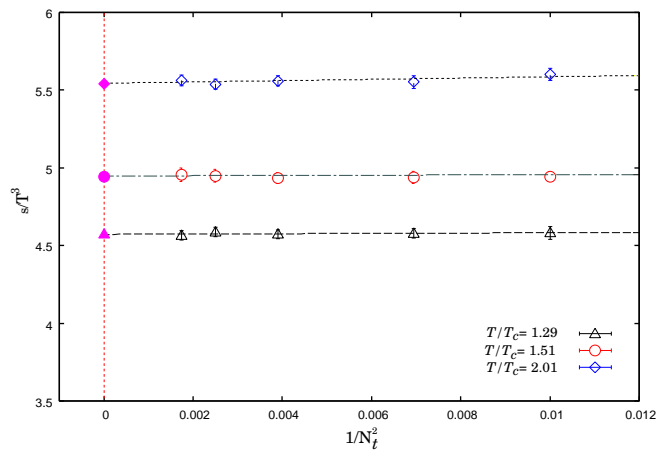}} 
\caption{\label{fig3} Entropy density $s/T^{3}$ as a function of $1/N_{t}^{2}$ at $T = 1.29, 1.51, 2.07 T_{c}$ together with continuum extrapolation fits using Eq. (\ref{eqnf1}). The solid symbols represent the results in the continuum limit.}
\end{figure}

By changing $N_{t}$ at fixed $\beta$, we observe the expected scaling of relative error with $N_{t}^{4}$. Fig. \ref{fig3} displays the continuum limit of entropy density at temperatures $T= 1.29, 1.51, 2.07T_{c}$ for $\xi = (1,0,0)$.  We fit the data to linear fits  $1/N_{t}^{2}$  and observe that the extrapolations provide good fits to the data. To gain an idea of the magnitude of finite volume effects in the entropy density, we consider the ratio between the entropy densities for  $N_{s}= 64$ and $128$ at various temperatures. We find that the ratio $s(N_{s}=128/s(N_{s}=64$ ranges between 0.986 to 0.992 for the temperatures $T= 1.51, 2.079, 2.491, 2.932T_{c}$.  This implies that the finite volume effects are less than 1.4$\%$. We note that the difference between the extrapolated values and the continuum results is just over a percent.

\begin{table}[ht!]
\begin{center}
\caption{Results of the continuum value of entropy density for $\pmb{\xi}= (1,0,0)$. The errors quoted are statistical.}
\label{tab3}
\begin{tabular}{ccccccccccc}\hline\hline
$T/T_{c}$ \hspace{1.0cm} &&&&&&& $s/T^{3}$ \\ \hline
1.064 \hspace{1.0cm} & &&&&&& 3.605(13) \\
1.123 \hspace{1.0cm} & &&&&&& 3.854(15) \\ 
1.294 \hspace{1.0cm} & &&&&&& 4.573(14) \\ 
1.513 \hspace{1.0cm} & &&& &&& 4.993(16) \\ 
1.860 \hspace{1.0cm} & &&& &&& 5.398(12) \\ 
2.079 \hspace{1.0cm} & &&& &&& 5.588(21) \\ 
2.491 \hspace{1.0cm} & &&& &&& 5.856(18) \\ 
2.784 \hspace{1.0cm} & &&& &&& 5.934(16) \\
2.932 \hspace{1.0cm} & &&& &&& 6.028(15) \\
3.051 \hspace{1.0cm} & &&& &&&6.068(20) \\ \hline
\end{tabular}
\end{center}
\end{table}

\begin{figure}[ht!]
\scalebox{0.45}{\includegraphics{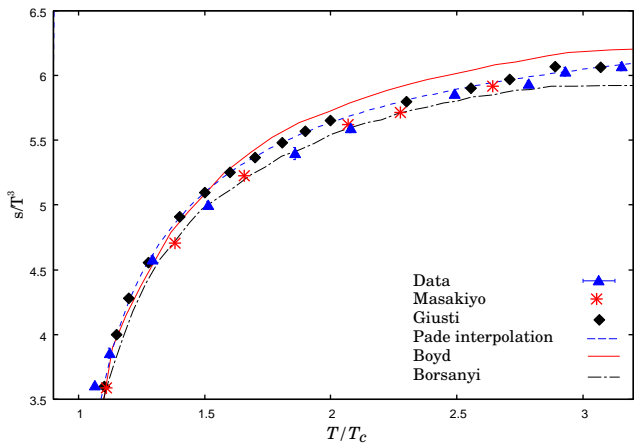}} 
\caption{\label{fig4} Temperature dependence of entropy density $s/T^{3}$ in the continuum limit. The solid red line shows the results from \cite{Boyd1996}, and the dashed lines (black and blue) represent results from \cite{Borsanyi2012} and the Pade interpolation formula for $T/T_{c}\in [1.0, 3.4]$. The red star symbols are the estimates using the gradient flow method \cite{Kitazawa2016}.} 
\end{figure}

The temperature dependence of extrapolated results of the entropy density is depicted in Fig. \ref{fig4} and summarised in Table \ref{tab3}. We explore the data sets in the region where the temperature dependence of the measured observables is stronger. As expected, the temperature dependence of the ratio $s/T^{3}$ shows a rapid increase in the region of the phase transition. This is followed by a relatively slow rise in the 2 - 3$T_{c}$ region. This suggests that entropy density, among other thermodynamical quantities, may only have logarithmic dependence on the temperature.
 We plot our results together with those obtained in Refs. \cite{Boyd1996,Borsanyi2012,Kitazawa2016} and the model predictions of Pade interpolating formula \cite{Giusti2017}. Above a few $T_{c}$ and within statistical errors, it can be seen that our continuum extrapolated data shows a good agreement with the results obtained using the modified integral method \cite{Borsanyi2012} and gradient flow approach \cite{Kitazawa2016} in the vicinity of the critical temperature in the deconfined region. 
 
We observe a disagreement with a discrepancy corresponding to a 3 - 6 percent effect with the results in Ref. \cite{Boyd1996} for $T\approx 1.5 - 3.07T_{c}$.  Whereas the primary observable in the approaches used in Refs. \cite{Boyd1996} and \cite{Borsanyi2012} is the interaction measure from which all other thermodynamic observables are calculated; the results obtained in these studies show significant differences in the region just above $T_{c}$. The ratio $s/T^{3}$  differs by approximately $4 - 6\%$  between 1.5$T_{c}$ and 3.4$T_{c}$. Such a disagreement close to the peak of the interaction measure has also been reported in other lattice studies \cite{Umeda2014}.  This could be due to the non-perturbative contribution contained in the trace anomaly, which dominates for $T_{c}<T<5T_{c}$ and reduces at increasing temperature. This nonperturbative contribution is quantified in the results for the trace anomaly in Ref. \cite{Borsanyi2012}. Our continuum values, however, compare well with the results in Ref. \cite{Borsanyi2012}. We report a similar agreement with the data (star symbols in Fig. \ref{fig4}) obtained in Ref. \cite{Kitazawa2016} using the gradient flow approach. The entropy density computed shows about $10\%$ deviation from the free theory value above $2.70T_{c}$.

The speed of sound is computed by differentiating the curve resulting from the fit of $\langle T_{0k}\rangle_{\xi}$. For $T>T_{c}$, the differential is calculated between each of consecutive points at which $\langle T_{0k}\rangle_{\xi}$ is measured. A quadratic fit in $\ln (T/T_{c})$  is used to interpolate $\langle T_{0k}\rangle_{\xi}$  between any two such points and the coefficients of the fits fixed by four data points close to the region of differentiation. The chi-square fit goodness test gauges the fit quality. The statistical errors on $dS/dT$ are computed by propagating linearly on $\langle T_{0k}\rangle_{\xi}$.

\begin{figure}[ht!]
\scalebox{0.45}{\includegraphics{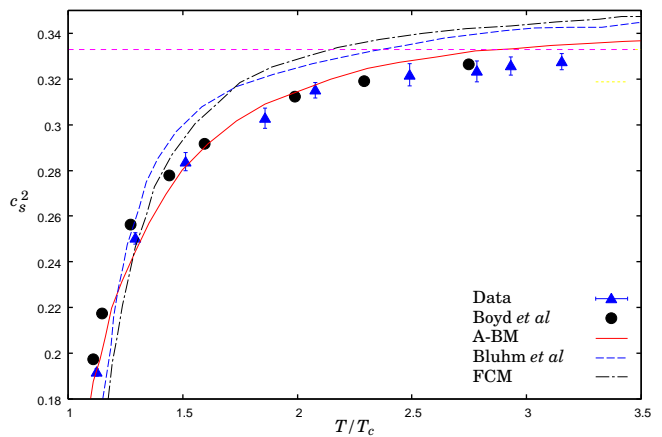}} 
\caption{\label{fig5} Speed of sound squared $c_{s}^{2}$ for SU(3) gluedynamics in comparison with the analytic predictions obtained from the modified bag model \cite{Begun2011} (solid red line), the field correlator method (FCM)  \cite{Khaidukov2018} (black dash-dot line) together with earlier numerical results of Refs. \cite{Boyd1996} (solid circles) and \cite{Bluhm2011} (blue dashed line). The dashed magenta line represents the Stephan-Boltzmann limit of the speed of sound for an ideal gas.}
\end{figure}

The results for the speed of sound squared, $c_{s}^{2}$, are plotted in Fig. \ref{fig5} in comparison with the results from the modified bag model \cite{Begun2011}, field correlator method (FCM) \cite{Khaidukov2018}, and previous lattice data \cite{Boyd1996, Bluhm2011}.  Near the critical temperature in the deconfining region, we observe a rapid decrease of $c_{s}^{2}$ in the vicinity of the critical temperature in the deconfinement region. The speed of sounds shows constancy at the temperatures $T> 2.5T_{c}$. Our results almost follow the modified bag model predictions and display consistency with the earlier lattice data \cite{Boyd1996} at higher temperatures.  The speed of sound is found in the region $c_{s}^{2} \leq 0.333$ for the range of temperatures $1.06T_{c}\leq T \leq 3.05T_{c}$. It can be seen that within the field correlator method using the nonperturbative colour magnetic confinement and Polyakov loop interaction in the deconfined region, and lattice results of Bluhm et al. \cite{Bluhm2011},   $\Delta = c_{s}^{2}-0.333$ is positive for $T> 2.32$. A similar behaviour is displayed by the modified bag model predictions for $T> 3T_{c}$. Since the scaling symmetry is significantly violated in the confinement-deconfinement transition region, the speed of sound is expected to deviate from that of an ideal gas of massless particles. It has been observed that the deviation from conformality is quite significant even at temperatures about $T= 500$ MeV. It has been suggested that the Generalised Uncertainty Principle (GUP) introduces a scale that breaks the prior conformal invariance of the system of noninteracting massless particles at higher $T$ \cite{Nagger2013,Elmashad2014,Salem2015,Nasser2018}, which hints that the lattice study of the QGP medium should be done with more care.

\section{Conclusions}

We have determined the entropy density of the pure SU(3) theory using $O(a^{4}$) improved lattice energy-momentum and field strength tensors. The numerical simulations were performed to study the temperature dependence of the entropy density near $T_{c}$. In the framework of shifted boundary conditions, the space-time matrix elements of the energy-momentum tensor have a nonvanishing expectation value related to the entropy density of the system by a purely multiplicative factor. Additionally, unlike the methods based on the measurement of the trace anomaly, this approach does not require the subtraction of ultraviolet power divergence. Once the renormalisation constant is known, one can straightway obtain the entropy density by computing the expectation value of space-time components of EMT.  The approach provides a more straightforward way to obtain the continuum limit of the entropy density of the system. 

The calculations performed with the improved discretisation on larger volumes have shown that finite volume effects are negligible. From the simulations on $64\times (8, 10, 12,16,20)$ lattices and relatively high statistics, the continuum extrapolation of the lattice data for entropy density was obtained with a few percent precision, including statistical and systematic errors. We find that at temperatures of about $3T_{c}$, deviations of entropy density from the Stefan-Boltzmann limit ($s_{SB}/T^{3}$)  are about 10$\%$. The slow approach to this limit agrees with the expectation that the functional dependence of thermodynamic observables in this regime is controlled by a running coupling that varies with the temperature only logarithmically. Our results agree well with the previous results obtained using the gradient flow method in the temperature region investigated in this study.  In the case of the improved discretisation, the magnitude of the $O(a^{4})$ corrections have been reduced strongly compared to the results obtained using one-plaquette action. The speed of sound is observed to be well-behaved near the critical temperature in the deconfined region and in good agreement with the results obtained in earlier lattice calculations. 

\section{Acknowledgements}
Numerical simulations for this study were carried out on the Shaheen III Supercomputer at the KAUST under its HPC Program.  We thankfully acknowledge the computer resources provided by KAUST.


\begin{references}
\bibitem{Sayantan2021}
S. Sharma,   Int. J. Mod. Phys. E {\bf 30}, 2130003 (2021)

\bibitem{Bernhard2016}
J. E. Bernhard, J. S. Moreland, S. A. Bass, J. Liu, and U. Heinz, Phys. Rev. C {\bf 94}, 024907 (2016)

\bibitem{Parotto2020}
P. Parotto \emph{et al}., Phys. Rev. C {\bf 101}, 034901 (2020)

\bibitem{Monnai2019}
A. Monnai, B. Schenke, and C. Shen, Phys. Rev. C {\bf 100}, 024907 (2019)

\bibitem{Everett2020}
D. Everett \emph{et al}., (JETSCAPE), (2020)
\bibitem{Derradi2016}
R. Derradi de Souza, T. Koide, and T. Kodama, Prog. Part. Nucl. Phys. 86, {\bf 35} (2016)

\bibitem{Molnar2002}
D. Molnar and M. Gyulassy, Nucl. Phys. A{\bf697}, 495 (2002); Erratum in Nucl. Phys. A{\bf703}, 893 (2002)

\bibitem{Gyu2005}
M. Gyulassy and L. McLerran, Nucl. Phys. {\bf A750}, 30 (2005)

\bibitem{Shuryak2005}
E. Shuryak, Nucl. Phys. {\bf A750}, 64 (2005)

\bibitem{Peshier2005}
A. Peshier and W. Cassing, Phys. Rev. Lett. {\bf 94}, 172301 (2005)

\bibitem{Heinz2013}
U. Heinz and R. Snellings, Annu. Rev. Nucl. Part. Sci. {\bf 63}, 12 (2013)

\bibitem{Oll2008}
J. Y. Ollitrault, Eur. J. Phys. {\bf 29}, 275 (2008) 

\bibitem{Gale2013}
C. Gale, S. Jeon and B. Schenke, Int. J. Mod. Phys. A {\bf 28}, 1340011 (2013)

\bibitem{Boyd1996}
G. Boyd \emph{et al}.,  Nucl. Phys. {\bf B 469}, 419 (1996)

\bibitem{Okamoto2001}
M. Okamoto \emph{et al}., [CP-PACS Collaboration], Phys. Rev. D {\bf 60}, 074507 (2001)

\bibitem{Umeda2009}
T. Umeda, S. Ejiri, S. Aoki, T. Hatsuda, K. Kanaya, Y. Maezawa and H. Ohno, Phys. Rev. D {\bf 79}, 051501 (2009)

\bibitem{Borsanyi2012}
S. Borsanyi, G. Endrodi, Z. Fodor, S. D. Katz and K. Szabo, JHEP {\bf 1207}, 056 (2012)

\bibitem{Giusti2016}
L. Giusti and M. Pepe, PoS LATTICE {\bf 2015}, 211 (2016)

\bibitem{Borsanyi2014}
S. Borsanyi, Z. Fodor, C. Hoelbling, S. D. Katz, S. Krieg and K. K. Szabo, Phys. Lett. B {\bf 730}, 99 (2014)

\bibitem{Bazavov2014}
A. Bazavov \emph{et al.},  [HotQCD Collaboration], Phys. Rev. D {\bf 90}, 094503 (2014) 

\bibitem{Philpsen2013}
O. Philipsen, Prog. Part. Nucl. Phys. {\bf 70}, 55 (2013)

\bibitem{Ding2015}
H.-T. Ding, F. Karsch and S. Mukherjee, Int. J. Mod. Phys. E {\bf 24}, 1530007 (2015) 1530007

\bibitem{Bazavov2018}
A. Bazavov, P. Petreczky and J. Weber, Phys. Rev D {\bf 97}, 014510 (2018)

\bibitem{Suzuki2013}
H. Suzuki, PTEP, {\bf 8}, 083B03 (2013); Erratum: PTEP, {\bf 7}, 079201 (2015)

\bibitem{Asakawa2015}
M. Asakawa \emph{et al}.,  [FlowQCD Collaboration], Phys. Rev. D {\bf 90}, 011501 (2014); Erratum: Phys. Rev. D {\bf 92}, 059902 (2015) 

\bibitem{Luscher2010}
M. Luscher, JHEP {\bf 1008}, 071 (2010) 

\bibitem{Narayanan2006}
R. Narayanan and H. Neuberger, JHEP {\bf 0603}, 064 (2006)

\bibitem{Fodor2012}
Z. Fodor, K. Holland, J. Kuti, D. Nogradi and C. H. Wong, JHEP {\bf 1211}, 007 (2012)

\bibitem{Luscher2011}
M. Luscher and P. Weisz, JHEP, {\bf 1102}, 051 (2011)

\bibitem{Makino2014}
H. Makino and H. Suzuki, PTEP, {\bf 6}, 063B02 (2014); Erratum: PTEP, {\bf 7}, 079202 (2015)

\bibitem{Itou2016}
E. Itou, H. Suzuki, Y. Taniguchi and T. Umeda, PoS LATTICE {\bf 2015}, 303 (2016) 

\bibitem{Taniguchi}
Y. Taniguchi, S. Ejiri, R. Iwami, K. Kanaya, M. Kitazawa, H. Suzuki, T. Umeda and N. Wakabayashi, arXiv:1609.01417 [hep-lat]

\bibitem{Giusti2011}
L. Giusti, H.B. Meyer, Phys. Rev. Lett. {\bf 106}, 131601 (2011) 

\bibitem{Giusti2011b}
L. Giusti, H.B. Meyer, J.High Energy Phys. {\bf 11}, 087 (2011) 

\bibitem{Giusti2013}
L. Giusti, H.B. Meyer, J.High Energy Phys. {\bf 1}, 140 (2013) 

\bibitem{Giusti2013M}
L. Giusti and H. Meyer, JHEP {\bf 1301}, 140 (2013)

\bibitem{Giusti2011M}
L. Giusti and H. Meyer, JHEP {\bf 1111}, 87 (2011);   JHEP {\bf 106}, 131601 (2011)

\bibitem{Giusti2017}
L. Giusti and M. Pepe, Phys. Lett. B {\bf 769}, 385 (2017)

\bibitem{Yagi2005}
K. Yagi, T. Hatsuda and Y. Miake, Camb. Monogr. Part. Phys. Nucl. Phys. Cosmol. {\bf 23}, 1 (2005)

\bibitem{Khan2006}
A. Khan et al., Phys. Rev. D {\bf 64}, 074510  (2001) 

\bibitem{Aoki2006}
Y. Aoki, Z Fodor, S. Katz and K. Szabo, J. High Energy Phys.  {\bf 01}, 089 (2006)

\bibitem{Alford1995}
M. Alford, W. Dimm, G. P. Lepage, G. Hockney, and P. B. Mackenzie, Nucl. Phys. B (Proc. Suppl.) {\bf 42}, 787 (1995); Phys. Lett. B {\bf 361}, 87 (1995).

\bibitem{Brida2018}
M. Dalla Brida, L. Giusti and M. Pepe, EJP Web Conf. {\bf 175}, 14012 (2018) 

\bibitem{Caracciolo1988}
S. Caracciolo, G. Curci, P. Menotti, and A. Pelissetto, Nucl.Phys. {\bf B309}, 612 (1988)

\bibitem{Kitazawa2016}
M. Kitazawa, T. Intani, M. Asakawa, T. Hatsuda and H. Suzuki, Phys. Rev. D {\bf 94}, 114512 (2016)

\bibitem{Umeda2014}
T. Umed, Phys. Rev. D {\bf 90}, 054511 (2014)

\bibitem{Begun2011}
V. Begun, M. Gorenstein, O. Mogilevsky, Int. J. Mod. Phys. E {\bf 20}, 1805, (2011)

\bibitem{Khaidukov2018}
Z. Khaidukov, M. Lukashov, Yu. Simonov, Phys. Rev. D {\bf 98}, 074031  (2018)

\bibitem{Bluhm2011}
M. Bluhm, B. Kampfer and K. Redlich, Phys. Rev. C {\bf 84}, 025201 (2011)

\bibitem{Nagger2013}
N. Naggar, L. Abou-Salem, I. Elmashad and A. Ali, J. Mod. Phys. {\bf 4}, 13 (2013)

\bibitem{Elmashad2014}
I. Elmashad, A. Ali, L. Abou-Salem, J. Nabi and A. Tawfik, SOP Trans. Theor. Phys. {\bf 1}, 1 (2014) 

\bibitem{Salem2015}
L. Abou-Salem, N. El Naggarand and I. Elmashad, Adv. High Energy Phys. {\bf 2015}, 103576 (2015) 

\bibitem{Nasser2018}
N. Demir and E. Vagenas, Nucl. Phys. B {\bf 933}, 340 (2018)

\end{references}
\end{document}